\documentclass[twocolumn,trackchanges]{aastex7}
\usepackage{amsmath}
\usepackage{calc}
\usepackage{enumitem}
\usepackage{multirow}
\usepackage{tabularx}
\usepackage{booktabs}
\usepackage{xcolor}

\begin{document}

\title{Inductive Biases in Field-Level Cosmological Inference from Galaxy Catalogs}

\author[orcid=0000-0002-3498-9086,sname='Baldwin']{James O. Baldwin}
\affiliation{City University of New York Graduate Center, 365 5th Ave, New York, NY 10016}
\email[show]{jbaldwin@gradcenter.cuny.edu}  

\author[orcid=0000-0002-3185-1540]{Shy Genel}
\affiliation{Center for Computational Astrophysics, Flatiron Institute, 162 5th Avenue, New York, NY 10010, USA}
\affiliation{Columbia Astrophysics Laboratory, Columbia University, 550 West 120th Street, New York, NY 10027, USA}
\email[]{sgenel@flatironinstitute.org}

\author[orcid=0000-0002-4816-0455, sname='Villaescusa-Navarro']{Francisco Villaescusa-Navarro}
\affiliation{Center for Computational Astrophysics, Flatiron Institute, 162 5th Avenue, New York, NY 10010, USA}
\affiliation{Department of Astrophysical Sciences, Princeton University, 4 Ivy Lane, Princeton, NJ 08544 USA}
\email[]{villaescusa.francisco@gmail.com}

\begin{abstract}

We perform field-level likelihood-free inference of the matter density parameter $\Omega_m$ from simulated galaxy catalogs using machine learning models with differing inductive biases. Using hydrodynamic simulations from CAMELS, we examine how observable choice and architecture govern cosmological information extraction. We consider galaxy positions and line-of-sight peculiar velocities, separately and jointly, and compare permutation-invariant Deep Sets, implemented with either multilayer perceptrons (MLPs) or Kolmogorov-Arnold Networks (KANs), to graph neural networks (GNNs), which explicitly encode spatial relations. We test in-distribution and out-of-distribution (OOD) performance across simulations with different subgrid galaxy-formation prescriptions. Deep Sets infer $\Omega_m$ from velocities alone with mean relative errors of approximately $18\%$ in-distribution and $\sim25\%$ OOD, with KANs and MLPs achieving comparable performance. In contrast, the same set-based approach does not yield useful $\sigma_8$ predictions in either in-distribution or cross-suite tests. Adding positions does not improve Deep Sets, while GNNs infer $\Omega_m$ with mean relative errors of about $10\%$ in-distribution and $10$--$17\%$ OOD. These results indicate that peculiar velocities provide the dominant source of $\Omega_m$ information for set-based models in this setting, while spatial information is most effectively used by architectures that explicitly encode galaxy-galaxy relations. Because the velocity inputs are exact simulated peculiar velocities, applications to survey data will require validation under realistic velocity-measurement noise, selection effects, and survey geometry.

\end{abstract}



\section{Introduction}\label{sec:intro}
The standard model of cosmology accurately describes a wide set of cosmological observations. As such, it is currently the most accepted theory for describing the evolution of our universe over the past 14 billion years. In this framework, dark matter (DM), baryonic matter, and dark energy (DE) fill the universe. The mysterious nature of DM and DE represents one of the greatest challenges currently facing cosmology, and understanding and constraining the parameters of the standard model allows for insights into the fundamental physics which governs them.  

The spatial distribution of galaxies in the universe depends directly on the cosmological parameters that cosmologists hope to constrain, with the clustering of these objects containing important information about the specific values of these parameters. Accurate inference of cosmological parameters from galaxy surveys is therefore a central goal of modern cosmology. Existing and upcoming experiments will deliver galaxy catalogs of unprecedented size and complexity, motivating the development of inference methods capable of extracting maximal cosmological information from high-dimensional, heterogeneous data: DESI \citep{DESI_2014, Eisenstein_DESI_collab, desicollaboration2016desiexperimentisciencetargeting, DESI_collab_earlydata, desicollaboration2026datarelease1dark}, Euclid \citep{Laureijs_2011_Euclid_Definition, Amendola_2013_Euclid_Cosmology, Racca_2016_Euclid_Mission, Euclid_Collaboration_Castro_2022}, Prime Focus Spectrograph (PFS) \citep{Takada_2014_PFS}, J-PAS \citep{Benitez_2014_JPAS}, Square Kilometer Array (SKA) \citep{Taylor_Braun_1999_SKA}, Roman \citep{Spergel_2015_WFIRST_AFTA, roman_colab}, JWST \citep{Gardner2006, Gardner2023}. Traditionally, cosmological inference has relied on carefully constructed summary statistics, which compress the data into lower-dimensional representations. A wide variety of summary statistics have been explored, including correlation functions and higher momenta \citep{VillaescusaNavarro_2020_Quijote, Gatti2025}, peak counts \citep{Li_2019_peak_counts, Harnois2024}, wavelets \citep{Valogiannis_Dvorkin_2021_wavelets, Eickenberg_2022_wavelets}, and others \citep{Banerjee_2020, Bayer_2021, Naidoo_2022}. While effective, this compression can lead to information loss and may lead to suboptimal inference of cosmological parameters, particularly in the presence of complex baryonic effects and survey systematics.

Machine learning methods provide an alternative approach by enabling field-level inference directly from simulated or observed data, without the need for manually constructed summaries. Recent work has demonstrated that neural-network-based models can successfully infer cosmological parameters from large-scale structure data using, for example, convolutional neural networks \citep{Pan_Liu_2020, Villanueva_Domingo_2022, Yip_2025_topology_cosmo_params}, Bayesian neural networks \citep{Hortua_2023}, physics-informed neural networks \citep{Charnock2020, Simon2024}, and neural-network surrogate models \citep{DeRose_2022_NN_LSS_theory}. However, the performance and robustness of such methods depend not only on the training data but also on the choice of observables and the inductive biases encoded in the model architecture \citep{battaglia2018relationalinductivebiasesdeep}. Understanding how these factors interact is essential for machine learning–based inference to be reliably deployed in survey analysis pipelines.

The term field-level inference is used in the literature to describe several related but methodologically distinct approaches. In explicit Bayesian field-level analyses, one typically specifies a forward model and likelihood for the density or tracer field, often with the goal of sampling a posterior over cosmological parameters, initial conditions, density fields, and nuisance parameters \citep[e.g.,][]{Jasche_Wandelt_2013, Jasche_Lavaux_2019, Cabass_2019_EFTLikelihood}. In simulation-based inference, neural posterior, likelihood, or ratio estimators are trained on simulations to approximate the corresponding Bayesian objects without requiring an analytic likelihood \citep[e.g.,][]{Papamakarios_Murray_2016, Greenberg_2019_SNPE, Cranmer_2020_SBI}. The approach taken here is different from both of these cases. We use supervised neural estimators trained directly on simulated galaxy catalogs to predict cosmological parameters, together with an auxiliary heteroscedastic variance estimate, without constructing an explicit likelihood or posterior sampler. We therefore use ``field-level'' to indicate that the models operate directly on the galaxy catalog rather than on hand-crafted summary statistics, and ``likelihood-free'' to indicate that the mapping is learned from simulations without evaluating an analytic likelihood, consistent with usage in closely related galaxy-catalog inference studies \citep{Villanueva_Domingo_2022, de_Santi_2023, Lemos_2024, Lee_2024, FieldLevelReview_2025}. Recent explicit field-level analyses have also quantified the information content of galaxy clustering using perturbative forward models and likelihood-based inference, including comparisons to two- and three-point statistics and applications to biased tracers, BAO reconstruction, stochasticity modeling, and primordial non-Gaussianity \citep{Schmidt_2019_EFTForward, Schmidt_2020_EFTLikelihood, Nguyen_2024_FieldLevelInfo, Akitsu_2025_FieldLevelBispectrum, Babic_2026_BAOFieldLevel, Rubira_Schmidt_2026_NoiseField, Bayer_2026_BAOReconstruction, Andrews_2026_PNGFieldLevel}. Our analysis does not attempt to measure the total information available to an optimal field-level likelihood or to benchmark against those likelihood-based information-content bounds. Instead, the information-content claims made here are conditional on the catalog observables, finite CAMELS volumes, and neural architectures tested in this work.

In this work, we focus on the inference of the matter density parameter, $\Omega_m$, from simulated galaxy catalogs. With the inference problem as our control, we probe the information content carried by different galaxy observables by using several machine learning architectures. Specifically, we investigate which features encode a cosmological signal that is accessible to permutation-invariant and graph-based models and which architectural assumptions are required to extract that signal effectively.

We primarily consider galaxy positions and line-of-sight peculiar velocities, both individually and in combination. Peculiar velocities are of particular interest because they are directly sourced by the large-scale gravitational potential and are therefore sensitive to cosmology, while being less directly tied to uncertain subgrid galaxy formation physics \citep{Howlett2017, Qin_2021, Abate2008, Turner2021, Turner2022, Turner2025, Tsagas2026}. At the same time, positional information is inherently relational, raising questions about whether certain models can fully exploit it without explicit spatial structure. 

The IEEE defines robustness in its standard glossary of software engineering terminology as ``the degree to which a system or component can function correctly in the presence of invalid inputs or stressful environmental conditions" \citep{ieee1990}. In machine learning, robustness specifically characterizes a model's ability to maintain predictive performance under a range of conditions and environmental changes \citep{braiek2024}. A central criterion in this work is cross-suite out-of-distribution (OOD) performance, defined here as generalization from one hydrodynamic simulation suite to galaxy catalogs generated with different hydrodynamic solvers and subgrid galaxy-formation prescriptions. This definition is restricted to cross-suite simulation transfer: we do not test distribution shifts in cosmological parameter ranges relative to those sampled by CAMELS, initial-condition statistics, redshift, simulation volume, survey geometry, selection functions beyond the adopted mass-cut marginalization, redshift-space distortions, or velocity-measurement errors. To assess this, we train each model on a single simulation suite and first evaluate performance on held-out data drawn from the same simulation. We then test the trained models on galaxy catalogs generated by independent simulations which share the same underlying cosmology but differ substantially in their implementations of baryonic feedback. Cross-simulation performance provides a controlled probe of which galaxy observables encode cosmological information that is stable under changes in subgrid physics, and which model architectures are able to extract that information without relying on simulation-specific patterns.

To explore these questions, we compare architectures that differ in how they handle the structure of galaxy catalogs. Deep Sets \citep{zaheer2018deepsets} enforce permutation invariance over variable-length inputs and can be instantiated with either multilayer perceptrons (MLPs) or Kolmogorov–Arnold Networks (KANs) \citep{liu2025kankolmogorovarnoldnetworks} as their universal function approximators. Graph neural networks (GNNs), which treat galaxies as nodes and connect them based on a linking radius, explicitly encode the relational structure through local proximity \citep{Gori, gilmer2017neuralmessagepassingquantum}. KANs have been proposed as an interpretable alternative to MLPs with favorable theoretical properties; we assess whether these advantages carry over to cosmological inference. GNNs have already been shown to perform well at inference of $\Omega_m$ when the input combines spatial information with informative galaxy features \citep{Villanueva_Domingo_2022, de_Santi_2023}. The main new contributions of this work are the comparison between KAN- and MLP-based Deep Set models, the systematic cross-suite evaluation across multiple CAMELS hydrodynamic models, and the identification of a velocity-only $\Omega_m$ signal in set-based architectures. The GNN experiments serve as a relational-architecture benchmark that confirms and contextualizes previous graph-based results.

Our aim is not to introduce a new inference method, but to clarify where cosmological information resides in galaxy catalogs and how different inductive biases govern a model’s ability to access it. By holding the inference task fixed and systematically varying both the feature set and the architecture, we identify which combinations are sufficient for robust inference and which require more structured models. The results have direct implications for the design of machine learning pipelines for upcoming galaxy surveys, as well as for the interpretation of machine learning-based cosmological constraints, provided realistic velocity-measurement noise, selection effects, and survey geometry are all accounted for.

\section{Simulations}
\label{sec:simulations}

We make use of first-generation simulations from the Cosmology and Astrophysics with MachinE Learning Simulations (\textsc{camels}) project~\citep{CAMELS_presentation, CAMELS_DR1, CAMELS_DR2}, a suite of thousands of state-of-the-art cosmological hydrodynamic and $N$-body simulations designed to train and test machine-learning algorithms across a wide range of cosmological and astrophysical conditions. All CAMELS hydrodynamic simulations considered here use periodic boxes of comoving volume $(25\,h^{-1}{\rm Mpc})^3$ with $256^3$ dark matter particles and initially $256^3$ gas resolution elements. The dark matter particle mass is $6.49\times10^7 (\Omega_m-\Omega_b)/0.251\,h^{-1}M_\odot$, while the initial gas mass resolution is $1.27\times10^7 (\Omega_b/0.049)$ $h^{-1}M_\odot$ \citep{CAMELS_presentation}. The finite mass and spatial resolution of CAMELS does not allow scales below $\sim 1\,h^{-1}{\rm kpc}$ to be reliably resolved.

CAMELS simulations are organized into different simulation \textit{suites}, each run with different hydrodynamic solvers, and various simulation \textit{sets}, which employ a range of schemes to sample different cosmological and astrophysical parameters \citep{CAMELS_presentation}. 

\subsection{Simulation Suites}
\label{subsec:suites}

We draw on four galaxy-formation models, each with a different hydrodynamics solver and subgrid physics implementation:

\begin{description}[leftmargin=2em, labelindent=2em]
    \item[\textbf{Astrid}] Run with \textsc{mp-gadget}~\citep{yu_feng_2018_1451799} and implementing a modified version of the subgrid physics used in the Astrid simulations \citep{Bird_2022, Ni2022, Ni2023}. We use the LH set (1,000 simulations).

    \item[\textbf{IllustrisTNG}] Run with \textsc{arepo}~\citep{Springel_2010, Weinberger2020} and utilizing the same subgrid physics as the IllustrisTNG simulations \citep{Weinberger_2016, Pillepich2018}. We use both the LH (1,000 simulations) and SB28 (2,048 simulations) sets.

    \item[\textbf{Simba}] Run with \textsc{gizmo}~\citep{Hopkins2015} and a modified version of the subgrid physics used in the original Simba simulations \citep{Dave2019, CAMELS_presentation}. We use the LH set (1,000 simulations).

    \item[\textbf{Swift-EAGLE}] A \textsc{camels} adaptation of the EAGLE model~\citep{Schaye2015} implemented in the \textsc{swift} code~\citep{Schaller2016, Schaller2018, Schaller2024} with the \textsc{sphenix} scheme~\citep{Borrow2021}, calibrated at \textsc{camels} resolution (Lovell et al. \textit{in prep}). We use the LH set (1,000 simulations).
\end{description}

We emphasize that although the four astrophysical parameters share the same labels across suites, they encode physically distinct processes in each model and should not be compared directly~\citep{CAMELS_presentation}. Table~\ref{tab:simulations} summarizes the simulation sets used in this work.
 
\begin{table*}
  \centering
  \begin{tabularx}{0.64\textwidth}{llllcc}
    \hline\hline
    Suite & Code & Hydro solver & Set & $N_{\rm sim}$ & $N_{\rm params}$ \\
    \hline
    \textsc{IllustrisTNG} & \textsc{arepo}     & Voronoi moving mesh & LH   & 1000 & 6  \\
    \textsc{IllustrisTNG} & \textsc{arepo}     & Voronoi moving mesh & SB28 & 2048 & 28 \\
    \textsc{Astrid}       & \textsc{mp-gadget} & SPH                  & LH  & 1000 & 6  \\
    \textsc{Simba}        & \textsc{gizmo}     & Meshless finite mass & LH  & 1000 & 6  \\
    \textsc{Swift-EAGLE}  & \textsc{swift}     & SPH (SPHENIX)        & LH  & 1000 & 6  \\
    \hline\hline
  \end{tabularx}
\caption{Summary of \textsc{camels} simulation sets used
in this work. All runs have a box size of $(25\,h^{-1}\,\mathrm{Mpc})^{3}$ and $256^3$ dark-matter and $256^3$ gas particles. ``LH'' denotes the Latin Hypercube set; ``SB28'' denotes the 28-dimensional Sobol sequence set. $N_{\rm sim}$ is the number of hydrodynamic simulations used from each set.}
\label{tab:simulations}
\end{table*}

\subsection{Simulation Sets}
\label{subsec:sets}

\textsc{camels} organizes simulations into \emph{sets} that differ in how parameter space is sampled. We use two sets in this work:

\begin{description}[leftmargin=2em, labelindent=2em]
    \item[\textbf{Latin Hypercube (LH)}] Contains 1,000 simulations per suite, each with a unique combination of all six free parameters arranged in a Latin hypercube, with a different initial random seed per simulation~\citep{Villaescusa-Navarro_2022}. The parameter ranges are $\Omega_{\rm m}\in[0.1,\,0.5]$, $\sigma_{8}\in[0.6,\,1.0]$, $A_{\rm SN1},\,A_{\rm AGN1}\in[0.25,\,4.0]$, and $A_{\rm SN2},\,A_{\rm AGN2}\in[0.5,\,2.0]$~\citep{CAMELS_presentation}.

    Across hydrodynamic suites, the LH sets sample the same parameter ranges but are independently generated, so catalogs with the same LH index do not correspond to matched cosmological/astrophysical parameter vectors or shared initial conditions.

    \item[\textbf{Sobol Sequence (SB28)}] Contains 2,048 simulations sampling a 28-dimensional parameter space via a Sobol quasi-random sequence \citep{SOBOL196786}, offering superior space-filling properties in high dimensions~\citep{Ni2023}. This set is unique to the \textsc{IllustrisTNG} suite. Note that for SB28, $\Omega_b$ is sampled linearly over $[0.029,0.069]$, so both the dark matter particle mass and initial gas mass resolution vary across the set.
\end{description}

\subsection{Galaxy Catalogs} \label{sec:catalogs}

All galaxy catalogs used in this work employ the \textsc{SubFind} method to identify halos and subhalos within the simulations at each snapshot \citep{10.1046/j.1365-8711.2001.04912.x, 10.1111/j.1365-2966.2009.15034.x}. The definition of a galaxy in the context of our simulations is any subhalo containing at least one star particle. In this work, following the procedure laid out in \cite{de_Santi_2023}, we only consider galaxies with stellar masses above $1.3\times10^8 M_{\odot}/h$, corresponding to $\sim10$ star particles in the LH simulations, and $\sim7$-17 star particles across the SB set due to the variation in $\Omega_b$. For each simulation, we produce several galaxy catalogs by varying the stellar mass threshold and keeping only those galaxies exceeding that threshold. A more detailed overview of this process follows in Section~\ref{sec:method}.
         
\subsection{Input Features} \label{sec:inputs}

For this work, we consider two galaxy features that have been shown in previous work to contain robust cosmological information \citep{de_Santi_2023, Yuan_2024_SimBIG, Euclid_Clustering_Overview}: galaxy positions and the z-component of the peculiar velocities ($v_z$). In observational settings, peculiar velocities are not measured directly, but are inferred from spectroscopic redshifts together with redshift-independent distance indicators such as Tully–Fisher \citep{TullyFisher1977}, the Fundamental Plane \citep{Djorgovski1987}, Faber–Jackson \citep{FaberJackson1976}, or Type Ia supernovae \citep{Branch1993}. Here we use the corresponding simulated quantity, though it should be noted that no corrections for observational or instrumentation error have been implemented.

To ensure the stability of the model during training, the peculiar velocities are transformed according to 

\begin{equation}
    v_z\mapsto \text{sign}(v_z)\cdot \log_{10}[1+abs(v_z)].
\end{equation}

For all experiments that include positional information, both Deep Sets and GNNs receive scalar position features rather than raw Cartesian coordinates. We construct these features relative to the fixed center of the simulation box,
\(\mathbf c=(L/2,L/2,L/2)\), with \(L=25\,h^{-1}{\rm Mpc}\). For each galaxy position \(\mathbf r_i\), we transform
\begin{equation}
    (x_i,y_i,z_i) \mapsto (d_i,\alpha_i,\beta_i),
\end{equation}
where
\begin{align}
    \boldsymbol{\delta}_i &= \mathbf r_i-\mathbf c, \\
    d_i &= |\boldsymbol{\delta}_i|, \\
    \alpha_i &=
    \arccos\left(
    \frac{\mathbf r_i\cdot \mathbf c}
    {|\mathbf r_i||\mathbf c|}
    \right), \\
    \beta_i &=
    \arccos\left(
    \frac{\boldsymbol{\delta}_i\cdot \mathbf c}
    {|\boldsymbol{\delta}_i||\mathbf c|}
    \right).
\end{align}
Thus, the position-only input associated with each galaxy is the three-dimensional scalar feature vector \((d_i,\alpha_i,\beta_i)\). When additional galaxy properties are used, such as \(v_z\), they are concatenated to this same pointwise feature vector. This parameterization removes the raw Cartesian coordinates from the network input and provides a fixed-dimensional scalar representation of each galaxy's position within the simulation box.

We also consider a global feature, the logarithm of the total number of galaxies in the simulation: $u=log(N)$\footnote{The option to include this feature was made a hyperparameter during optimization, and was found to slightly improve results.}. 

\section{Architectures}

\subsection{Deep Sets}

Deep Sets provide a neural architecture tailored to inputs that are unordered sets, making them particularly suited for modeling data where permutation invariance is essential \citep{zaheer2018deepsets}. In cosmological applications, each galaxy within a group contributes to the inference of global properties like the matter density parameter $\Omega_m$, but the order in which the galaxies are presented to the model is irrelevant. Traditional neural networks fail to respect this symmetry, whereas Deep Sets are explicitly constructed to do so.

The core theoretical result underpinning Deep Sets is that any permutation-invariant function $f$ operating on a set $X = \{x_1, x_2, \dots, x_n\}$ can be decomposed in the form:
\begin{equation}
    f(X) = \rho\left( \sum_{i=1}^n \phi(x_i) \right),
    \label{eq:deep_sets}
\end{equation}
where $\phi$ is a learned function that maps each element to a latent space, the summation acts as a permutation-invariant aggregation operator, and $\rho$ maps the aggregated latent representation to the final output. This result is analogous to the universal approximation theorem but extended to functions on sets \citep{zaheer2018deepsets}.

In practice, both $\phi$ and $\rho$ are typically implemented as multilayer perceptrons (MLPs). Given a set of galaxies, where each element $x_i \in \mathbb{R}^d$ represents the features, we compute an embedding $\phi(x_i) \in \mathbb{R}^k$, aggregate these embeddings across the set, and then process the result through $\rho$ to predict a global scalar like $\Omega_m$, along with uncertainties. This approach has multiple desirable properties in the context of cosmological inference: it allows for variable-length inputs, preserves invariance under permutations, and provides a natural architecture for parallelization across individual objects in a set. 

\subsection{The Multilayer Perceptron (MLP) as the Standard Function Approximator}

The multilayer perceptron (MLP) originated as a layered extension of the perceptron introduced by \citet{rosenblatt1958perceptron} and further elaborated in \citet{rosenblatt1962principles}. However, the limitations of single-layer perceptrons, most prominently highlighted by \citet{minsky69perceptrons}, implicitly motivated architectures incorporating hidden units. Early gradient-based learning rules emerged through the delta rule for adaptive linear neurons \citep{widrow1960asc}, while more general theoretical developments in gradient-based learning and reverse-mode differentiation established in principle how multilayer networks could be trained by gradient descent (\cite{Amari1967ATO}: early treatments of adaptive gradient methods for pattern classifiers, \cite{10.1007/BF01931367}: formal development of reverse-mode automatic differentiation and \cite{Werbos:74_beyond}: its explicit application to neural network training). \citet{rumelhart1986learning} subsequently demonstrated that backpropagation could efficiently train multilayer networks in practice, establishing the modern multilayer perceptron as a viable learning architecture.

In its modern form, an MLP consists of a sequence of fully connected layers that transform a fixed-dimensional input vector through learnable affine mappings followed by nonlinear activation functions (e.g. ReLU, SiLU, or $\tanh$). The network parameters are optimized via gradient descent using the backpropagation algorithm. Owing to the universal approximation theorem \citep{Cybenko1989,HORNIK1989359}, MLPs serve as general-purpose function approximators and remain a foundational building block of many neural network architectures. 

\subsection{Kolmogorov–Arnold Networks (KANs) and DeepKANs}

Kolmogorov–Arnold Networks (KANs) \citep{liu2024kan20kolmogorovarnoldnetworks, liu2025kankolmogorovarnoldnetworks} are neural architectures motivated by the Kolmogorov–Arnold representation theorem (KART), which states that any continuous multivariate function defined on a bounded domain can be expressed as a finite composition of continuous univariate functions and addition \citep{Kolmogorov1957}. In a standard multi-layer perceptron (MLP), each node applies a fixed nonlinear activation function after a linear transformation along the edges, whereas in a KAN the nonlinear transformations are learned functions on the edges rather than at the nodes. Each edge carries a learnable one-dimensional function, while nodes perform summation. This structural inversion fundamentally alters how function approximation is realized within the network.

A depth-$L$ KAN is specified by a width array $[n_0, n_1, \dots, n_L]$, where $n_\ell$ denotes the number of neurons in layer $\ell$. A KAN layer $\Phi_\ell$ maps $x_\ell \in \mathbb{R}^{n_\ell}$ to $x_{\ell+1} \in \mathbb{R}^{n_{\ell+1}}$ through a matrix of univariate functions $\{\phi_{\ell,q,p}\}$, where each $\phi_{\ell,q,p} : \mathbb{R} \rightarrow \mathbb{R}$ acts on a single scalar input. The output of neuron $q$ in layer $\ell+1$ is given by
\begin{equation}
x_{\ell+1}^{(q)} = \sum_{p=1}^{n_\ell} \phi_{\ell,q,p}\!\left(x_\ell^{(p)}\right).
\end{equation}
The full network is the composition of such layers, $\mathrm{KAN}(x) = (\Phi_{L-1} \circ \cdots \circ \Phi_0)(x)$, producing a global multivariate function assembled from learned univariate transformations and summation.

Each edge function $\phi(x)$ is parameterized as a residual expansion of a B-spline. Specifically,
\begin{equation}
\phi(x) = w_b\, b(x) + w_s \sum_{i=0}^{G+k-1} c_i B_i(x),
\end{equation}
where $b(x)$ is a fixed residual basis function (typically SiLU), $\{B_i(x)\}$ are $k$-th order B-spline basis functions defined on a grid with $G$ intervals, $c_i$ are trainable spline coefficients, and $w_b$ and $w_s$ are learnable scaling parameters. The spline basis is constructed over a knot vector containing $G+1$ interior knots augmented by $k$ additional boundary knots, yielding $G+k$ basis functions. The grid parameter $G$ controls the resolution at which the univariate transformation is represented, providing a direct and interpretable knob for adjusting functional expressivity.

The KAN 2.0 framework extends this architecture by introducing multiplicative structure through Multiplicative KANs (MultKANs) \citep{liu2024kan20kolmogorovarnoldnetworks}. While standard KAN layers rely exclusively on addition at nodes, MultKAN layers incorporate explicit multiplication operators following the univariate transformation stage. Formally, a MultKAN layer can be written as $\Psi_\ell = M_\ell \circ \Phi_\ell$, where $\Phi_\ell$ is a standard KAN layer and $M_\ell$ applies designated element-wise products to selected outputs. The architecture is specified by additive and multiplicative width arrays, allowing controlled introduction of product interactions. When multiplicative nodes are absent, the MultKAN reduces to a standard KAN. This extension is motivated by the observation that many scientific relationships contain explicit multiplicative structure, and incorporating such operations directly can improve both interpretability and symbolic recoverability \citep{liu2024kan20kolmogorovarnoldnetworks}.

In the present work, KANs are embedded within a permutation-invariant Deep Set framework to form what we refer to as a DeepKAN. Recall that a Deep Set has the general form of Eq~\ref{eq:deep_sets}. In DeepKAN, both $\phi$ and $\rho$ are implemented as Kolmogorov-Arnold (or multiplicative Kolmogorov-Arnold) networks. The aggregation operation preserves permutation invariance over galaxies, while the spline-based edge parameterization provides flexible and interpretable functional transformations. The grid size $G$ thus controls the resolution of the learned univariate mappings applied at both the pointwise and post-aggregation stages. When multiplicative nodes are enabled, the architecture can explicitly represent product structures that may arise in relationships among galaxy observables.

From a scientific standpoint, DeepKANs provide a structured means of increasing local functional expressivity while preserving exchangeability with MLPs. The spline parameterization allows inspection of learned one-dimensional transformations and, in favorable cases, symbolic simplification of the resulting mappings. As shown in Section~\ref{sec:results}, however, increased local expressivity does not by itself lead to improved cosmological parameter inference relative to MLP-based Deep Sets in the case of inference from peculiar velocities. This suggests that, in this setting, architectural inductive bias and information content play a more decisive role than the specific form of univariate function approximation.

\subsection{Graph Neural Networks (GNNs) for Cosmological Inference} \label{sec:GNNs}

Graph neural networks (GNNs) are a class of machine learning architectures designed for data that can be represented as graphs, consisting of nodes, which contain information on associated features, and edges that encode relationships between nodes \citep{Gori, scarselli_gnns, Gallicchio, zhou2020gnn_review, wu2021gnn_survey}. Unlike set-based models, which treat elements independently prior to global aggregation, GNNs explicitly model interactions between neighboring elements through iterative message passing.

In a typical message-passing framework, each node updates its representation by aggregating information from its connected neighbors, allowing the network to learn relational structure \citep{gilmer2017neuralmessagepassingquantum}. After several such updates, node embeddings encode not only local features but also information about the surrounding environment. A global pooling operation can then be applied to obtain a graph-level representation useful for parameter inference. Explicitly, 

\begin{equation}
    \boldsymbol{h}_i^{k+1}=\phi^{(k)}\bigg(\boldsymbol{h}_i^{(k)},\bigoplus_{j \in \mathcal{N}(i)}\psi^{(k)}\big(\boldsymbol{h}_i^{(k)}, \boldsymbol{h}_j^{(k)} \big) \bigg),
\end{equation}

\noindent where $\phi^{(k)}$ and $\psi^{(k)}$ are learnable functions, and $\bigoplus$ represents a permutation-invariant aggregation. For the GNNs used in this work, both $\phi^{(k)}$ and $\psi^{(k)}$ are parameterized as MLPs. We forgo testing a GNN implementation using KANs based on their lack of improvement over MLPs in the set-based framework.

For cosmological galaxy catalogs, nodes correspond to galaxies and edges are typically constructed based on spatial proximity. This structure enables the model to learn correlations induced by gravitational clustering, tidal environments, and large-scale structure formation \citep{Villanueva_Domingo_2022, Jagvaral_2022_IA_GNN, DeSanti_2026_QGNN}. By incorporating relational inductive bias, GNNs are particularly well-suited for extracting information encoded in spatial configurations that may not be accessible to permutation-invariant architectures.

We adopt the GNN architecture used in \cite{Villanueva_Domingo_2022} in which the nodes contain galaxy feature information and the edges between galaxies are said to be connected if the radial distance between them is smaller than some parameter $r_{link}$.

\subsection{Performance Metrics}

To evaluate model performance, we employ a set of standard regression metrics that quantify both accuracy and statistical consistency between predicted and true values. We denote the true values as $y_i$, the model predictions as $\hat{y}_i$, and the predicted uncertainties as $\sigma_i$.

\begin{itemize}

\item \textbf{Root Mean Squared Error (RMSE):}
\begin{equation}
\mathrm{RMSE} = \sqrt{\frac{1}{N} \sum_{i=1}^{N} (\hat{y}_i - y_i)^2}.
\end{equation}
This metric provides an absolute measure of prediction error in the same units as the target. Lower values indicate higher precision.

\item \textbf{Coefficient of Determination ($R^2$):}
\begin{equation}
R^2 = 1 - \frac{\sum_{i=1}^{N} (y_i - \hat{y}_i)^2}{\sum_{i=1}^{N} (y_i - \bar{y})^2},
\end{equation}
where $\bar{y} = \frac{1}{N}\sum_{i=1}^{N} y_i$. This metric quantifies the fraction of variance explained by the model, with values approaching unity indicating high accuracy.

\item \textbf{Pearson Correlation Coefficient (PCC):}
\begin{equation}
\mathrm{PCC} = \frac{\mathrm{cov}(y, \hat{y})}{\sigma_y \sigma_{\hat{y}}}.
\end{equation}
This statistic measures the strength of the linear relationship between predictions and true values. Values close to $\pm 1$ indicate strong correlation, while values near $0$ indicate weak correlation.

\item \textbf{Bias:}
\begin{equation}
b = \frac{1}{N} \sum_{i=1}^{N} (\hat{y}_i - y_i).
\end{equation}
This metric captures systematic offsets between predictions and ground truth. Values close to zero indicate minimal bias.

\item \textbf{Mean Relative Error ($\epsilon$):}
\begin{equation}
\epsilon = \frac{1}{N} \sum_{i=1}^{N} \left| \frac{y_i - \hat{y}_i}{y_i} \right|.
\end{equation}
We report this quantity as a percentage. It provides a scale-independent measure of predictive accuracy.

\item \textbf{Reduced Chi-Squared ($\chi^2$):}
\begin{equation}
\chi^2 = \frac{1}{N} \sum_{i=1}^{N} \frac{(\hat{y}_i - y_i)^2}{\sigma_i^2}.
\end{equation}
This statistic evaluates the consistency between prediction errors and the model-predicted uncertainties. To avoid numerical instabilities, extreme outliers with $\chi_i^2 > 10^4$ are excluded prior to averaging.

\end{itemize}

To assess the calibration of the heteroscedastic uncertainty estimates, we compute empirical coverage as a function of the predicted uncertainty. Test catalogs are binned by predicted $\sigma_i$, and within each bin we measure the fraction of true $\Omega_m$ values lying within $\mu_i\pm\sigma_i$ and $\mu_i\pm2\sigma_i$. We also report aggregate coverage fractions over the full test set,

\begin{equation}
f_{n\sigma}=\frac{1}{N}\sum_{i=1}^{N}
\mathbf{1}\left(|\Omega_{m,i}^{\rm true}-\mu_i|\le n\sigma_i\right),
\qquad n=1,2,
\label{eq:coverage}
\end{equation}

where $\mu_i$ and $\sigma_i$ are the predicted mean and uncertainty for catalog $i$. For a calibrated Gaussian predictive distribution, $f_{1\sigma}$ and $f_{2\sigma}$ should approach the nominal values of $0.683$ and $0.954$, respectively. We show this diagnostic for representative in-distribution and cross-suite tests in Figure~\ref{fig:coverage}. Because $\chi^2$ averages squared standardized residuals, it is sensitive to the high-$\chi_i^2$ tail of the per-catalog residual distribution, even when aggregate coverage is close to nominal. We therefore interpret $\chi^2$ together with the empirical coverage fractions rather than as a stand-alone calibration diagnostic.

\section{Experimental Design} \label{sec:method}

For our training data, we begin with galaxy catalogs corresponding to the $z=0$ snapshot from the LH set. These simulations are separated into an 80/10/10 split between training, validation, and test sets. From the raw simulations, a marginalization over mass cuts is applied in which only galaxies above a specific mass cut are kept. Specifically, we apply cuts based on the prescription 

\begin{equation}
    M_{\star} > 1.3R\times10^8 M_{\odot}/h,
\end{equation}

\noindent where R is a uniformly distributed random value between 1 and 2. For each simulation in the set, we apply 10 mass cuts using this formula, resulting in 10,000 total simulations split among the training, validation, and test sets. This marginalization follows the procedure from \cite{de_Santi_2023}, representing a response to the results from \cite{Villanueva_Domingo_2022} whose models failed to generalize due to uniform mass cuts being unable to account for the variable galaxy populations resulting from different subgrid physics used in different simulations. For the test sets, the same procedure is followed for Simba, Swift-EAGLE, and the SB28 set of IllustrisTNG. We use Astrid as the primary training suite for two related reasons. First, this choice follows the velocity-based experiments of \citet{de_Santi_2023}, enabling a direct comparison with previous work and with the reproduction test presented in Appendix~\ref{appx:desanti}. Second, Astrid spans a broad range of galaxy catalog sizes across the LH set, broader than the corresponding ranges in the main IllustrisTNG and Simba LH suites, and therefore provides training support over much of the galaxy-count variation encountered in cross-suite tests \citep[Figure~\ref{fig:metalayer}]{de_Santi_2023}. This consideration is particularly relevant because the total number of galaxies is included as a global feature, $u=\log(N)$. Training on Astrid therefore reduces the possibility that cross-suite performance is dominated by a catalog-cardinality mismatch between the training and test sets, rather than by cosmological information carried by the input features.

All models are trained using a modified two-moment loss function adapted from \cite{jeffrey2020solvinghighdimensionalparameterinference}, designed to jointly learn parameter estimates and associated uncertainties. Specifically, the network predicts both a mean $\mu$ and variance $\sigma^2$, and is optimized using

\begin{equation}
\begin{split}
\mathcal{L}
&= \sum_i \log\!\left( \sum_{j \in \mathrm{batch}} (\theta_{i,j} - \mu_{i,j})^2 \right) \\
&\quad + \sum_i \log\!\left( \sum_{j \in \mathrm{batch}} \left[(\theta_{i,j} - \mu_{i,j})^2 - \sigma_{i,j}^2 \right]^2 \right)
\end{split}
\end{equation}

\noindent where $\theta_{i,j}$ denotes the true value of parameter $i$ for sample $j$. The first term enforces accurate mean predictions, while the second term encourages the predicted variance to match the squared residuals, enabling the model to learn heteroscedastic uncertainties. Following \cite{Villaescusa-Navarro_2022}, we replace the arithmetic sum used in the original formulation with a sum of logarithms, which empirically stabilizes training and yields more reliable estimates of both $\mu$ and $\sigma$. This modification prevents the loss from being dominated by parameters that are more easily constrained, allowing the network to balance contributions across parameters \citep{Villaescusa-Navarro_2022}.

The models are optimized using Optuna \citep{ozaki2025optunahub}, with hyperparameters set depending on the particular configuration. For the Deep Set with MLPs, the hyperparameters include the number and size of the hidden layer/s, whether to use residual layers, and whether to use the global parameter $u$\footnote{For all reported Deep Set and GNN results, the selected Optuna configurations used the global feature $u=\log N$.}. For Deep Sets with KANs, the hyperparameter list includes the number of addition and multiplication nodes, the intermediate KAN width \(n_\ell\) (which sets the dimensionality of the hidden representation passed from KAN layer \(\ell\) to layer \(\ell+1\)), the sizes of the two grids, whether to use a residual layer, and whether to use the global parameter $u$. The GNN from \cite{Villanueva_Domingo_2022}, similarly to the MLP configuration, optimizes the number and size of the hidden layers while additionally optimizing the length of $r_{link}$. All models optimize the learning rate and weight decay. 

\section{Results} \label{sec:results}
We first present the results of training Deep Sets with both MLPs and KANs using only the peculiar velocity information as input. We discuss the performance of these models and also present the results when including positional information along with peculiar velocities in the input. We then show how GNNs perform on our inference task, using both positional and peculiar velocity information as input. 

Table~\ref{tab:baselines} reports two simple baselines: predicting the Astrid training-set mean of $\Omega_m$, and linearly regressing $\Omega_m$ on the global catalog-size feature $u=\log N$. The $u=\log N$ baseline performs slightly better than the training-mean baseline, indicating that catalog abundance carries some cosmological information. However, both baselines remain substantially worse than the trained velocity-based models, whose RMSE values are typically $\sim0.055$--$0.081$ with mean relative errors of $\sim17\%$--$26\%$. Thus, the Deep Set models are not simply exploiting the training prior or the global number-count feature.

\subsection{Deep Sets: MLPs vs KANs using Peculiar Velocities}

\begin{figure*}
    \centering
    \includegraphics[width=\linewidth]{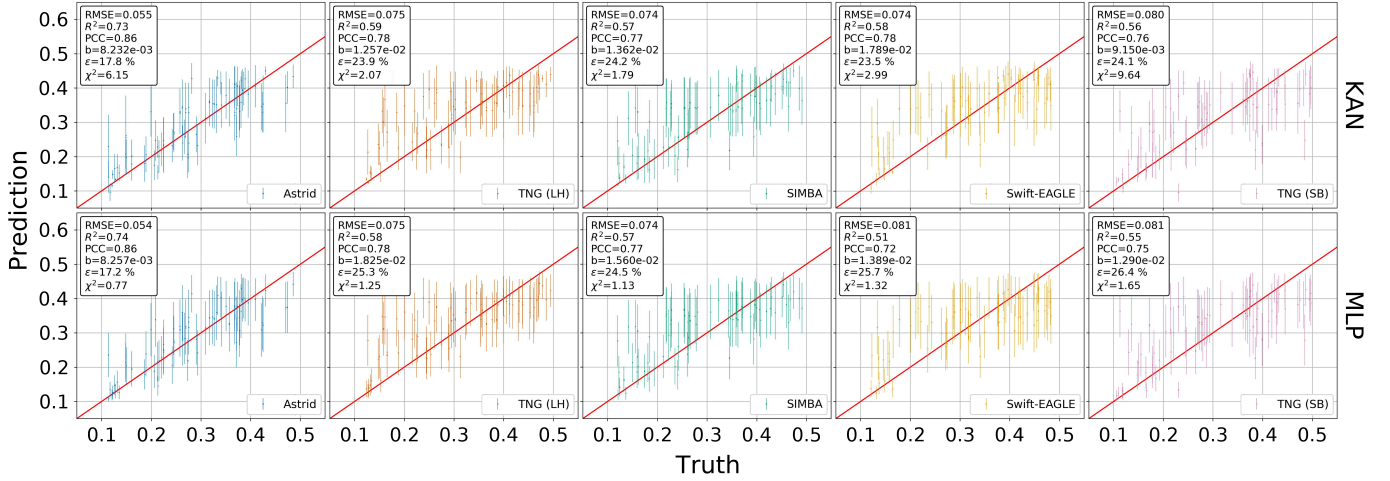}
    \caption{Comparing $\Omega_m$ inference using a DeepKAN model (top) and a Deep Set model implemented with multilayer perceptrons (MLPs) (bottom), trained on the LH set of the Astrid simulation suite, using only the line-of-sight peculiar velocities ($v_z$) as input. Performance on the training suite is comparable between the two models, as is the degradation in performance when tested on out-of-distribution data. This suggests that, in the velocity-only regime, performance is primarily limited by the information content of the input rather than by the choice of function approximator.}
    \label{fig:kan_vz_compare}
\end{figure*}

Figure~\ref{fig:kan_vz_compare} shows the performance of DeepKANs trained using only line-of-sight peculiar velocities and the analogous Deep Set model using standard MLPs. We find no substantial performance gain from replacing MLP-based function approximators with KANs in this setting. Both architectures perform comparably with respect to RMSE and $R^2$ on in-distribution data, and exhibit similar levels of cross-suite degradation when evaluated on TNG (both LH and SB sets), SIMBA, and Swift-EAGLE.

\begin{table}[t]
\centering
\begingroup  
\begin{tabular}{lcccc}
\toprule
\textbf{Test suite} 
& \multicolumn{2}{c}{\textbf{Mean}} 
& \multicolumn{2}{c}{$\log N$} \\
\cmidrule(lr){2-3}\cmidrule(lr){4-5}
& \textbf{RMSE} & \textbf{$\epsilon$ (\%)} 
& \textbf{RMSE} & \textbf{$\epsilon$ (\%)} \\
\midrule
ASTRID       & 0.107 & 45.0 & 0.105 & 43.3 \\
TNG (LH)     & 0.117 & 42.9 & 0.111 & 39.7 \\
TNG (SB)     & 0.121 & 48.1 & 0.117 & 44.3 \\
SIMBA        & 0.113 & 44.2 & 0.106 & 42.4 \\
Swift-EAGLE  & 0.117 & 41.5 & 0.099 & 36.1 \\
\bottomrule
\end{tabular}

\caption{Simple baselines for inferring $\Omega_m$. The mean baseline predicts the mean value of $\Omega_m$ in the ASTRID training split, $\langle\Omega_m\rangle_{\rm train}=0.296884$, for every catalog. The $\log N$ baseline fits a linear model, $\hat{\Omega}_m = a\log N + b$, using only the global catalog-size feature from the ASTRID training split, giving $a=0.0659$ and $b=0.1100$. Both baselines are evaluated on the same test suites as the trained models.}
\label{tab:baselines}
\endgroup
\end{table}

\begin{table*}[t]
\centering
\begin{tabularx}{0.8\textwidth}{llccccccc}
\hline
\textbf{Suite} & \textbf{UFA (features)} & \textbf{RMSE} & \textbf{$R^2$} & \textbf{PCC} & \textbf{Bias} & \textbf{$\epsilon$ (\%)} & \textbf{$\chi^2$} & \textbf{$N_{\rm clip}$} \\
\hline

\multirow{5}{*}{\textbf{ASTRID}}
& KAN ($v_z$)              & 0.055 & 0.73 & 0.86 & $8.23\times10^{-3}$ & 17.8 & 6.15 & 0 \\
& MLP ($v_z$)              & 0.054 & 0.74 & 0.86 & $8.26\times10^{-3}$ & 17.2 & 0.77 & 0 \\
& KAN (pos+$v_z$)          & 0.062 & 0.66 & 0.82 & $3.34\times10^{-3}$ & 19.4 & 1.76 & 0 \\
& MLP (pos+$v_z$)          & 0.054 & 0.74 & 0.86 & $1.85\times10^{-3}$ & 16.3 & 1.50 & 0 \\
& KAN (pos)                & 0.105 & 0.01 & 0.27 & $2.36\times10^{-2}$ & 43.6 & 1.21 & 0 \\
\hline

\multirow{5}{*}{\textbf{TNG (LH)}}
& KAN ($v_z$)              & 0.075 & 0.59 & 0.78 & $1.57\times10^{-2}$ & 23.9 & 2.07 & 0 \\
& MLP ($v_z$)              & 0.075 & 0.58 & 0.78 & $1.83\times10^{-2}$ & 25.3 & 1.25 & 0 \\
& KAN (pos+$v_z$)          & 0.067 & 0.67 & 0.82 & $5.84\times10^{-3}$ & 20.2 & 1.37 & 0 \\
& MLP (pos+$v_z$)          & 0.072 & 0.62 & 0.79 & $1.00\times10^{-2}$ & 22.9 & 2.06 & 0 \\
& KAN (pos)                & 0.114 & 0.03 & 0.20 & $-5.02\times10^{-3}$ & 42.3 & 1.09 & 0 \\
\hline

\multirow{5}{*}{\textbf{TNG (SB)}}
& KAN ($v_z$)              & 0.080 & 0.56 & 0.76 & $9.15\times10^{-3}$ & 24.1 & 9.64 & 0 \\
& MLP ($v_z$)              & 0.081 & 0.55 & 0.75 & $1.29\times10^{-2}$ & 26.4 & 1.65 & 0 \\
& KAN (pos+$v_z$)          & 0.068 & 0.68 & 0.83 & $-1.74\times10^{-4}$ & 20.9 & 1.55 & 0 \\
& MLP (pos+$v_z$)          & 0.080 & 0.56 & 0.75 & $1.12\times10^{-3}$ & 24.3 & 2.71 & 0 \\
& KAN (pos)                & 0.117 & 0.06 & 0.27 & $5.05\times10^{-3}$ & 46.8 & 1.15 & 0 \\
\hline

\multirow{5}{*}{\textbf{SIMBA}}
& KAN ($v_z$)              & 0.074 & 0.57 & 0.77 & $1.36\times10^{-2}$ & 24.2 & 1.79 & 1 \\
& MLP ($v_z$)              & 0.074 & 0.57 & 0.78 & $1.56\times10^{-2}$ & 24.5 & 1.13 & 0 \\
& KAN (pos+$v_z$)          & 0.077 & 0.53 & 0.74 & $1.03\times10^{-2}$ & 23.8 & 1.70 & 0 \\
& MLP (pos+$v_z$)          & 0.071 & 0.60 & 0.78 & $8.82\times10^{-3}$ & 21.8 & 1.89 & 0 \\
& KAN (pos)                & 0.109 & 0.07 & 0.35 & $1.50\times10^{-2}$ & 43.4 & 1.03 & 0 \\
\hline

\multirow{5}{*}{\textbf{Swift-EAGLE}}
& KAN ($v_z$)              & 0.074 & 0.58 & 0.78 & $1.78\times10^{-2}$ & 23.5 & 2.99 & 0 \\
& MLP ($v_z$)              & 0.081 & 0.51 & 0.72 & $1.39\times10^{-2}$ & 25.7 & 1.32 & 0 \\
& KAN (pos+$v_z$)          & 0.070 & 0.63 & 0.80 & $1.05\times10^{-3}$ & 20.3 & 1.42 & 0 \\
& MLP (pos+$v_z$)          & 0.079 & 0.53 & 0.73 & $5.87\times10^{-3}$ & 24.4 & 2.25 & 0 \\
& KAN (pos)                & 0.100 & 0.24 & 0.51 & $3.83\times10^{-3}$ & 36.8 & 0.95 & 0 \\
\hline

\end{tabularx}
\caption{Comparison of Deep Set models using different universal function approximators (UFAs) and different input features for inferring $\Omega_m$. All models are trained on ASTRID and evaluated on multiple simulation suites. We compare line-of-sight velocities ($v_z$) alone and in combination with positional information (pos+$v_z$). Here $\chi^2$ is computed after excluding catalogs with $\chi^2>10^4$, and $N_{\rm clip}$ gives the number of excluded catalogs. The global feature $u=\text{log}N$ is included in all training and test samples.}
\label{tab:feature_table}
\end{table*}

\begin{figure*}
    \centering
    \includegraphics[width=\linewidth]{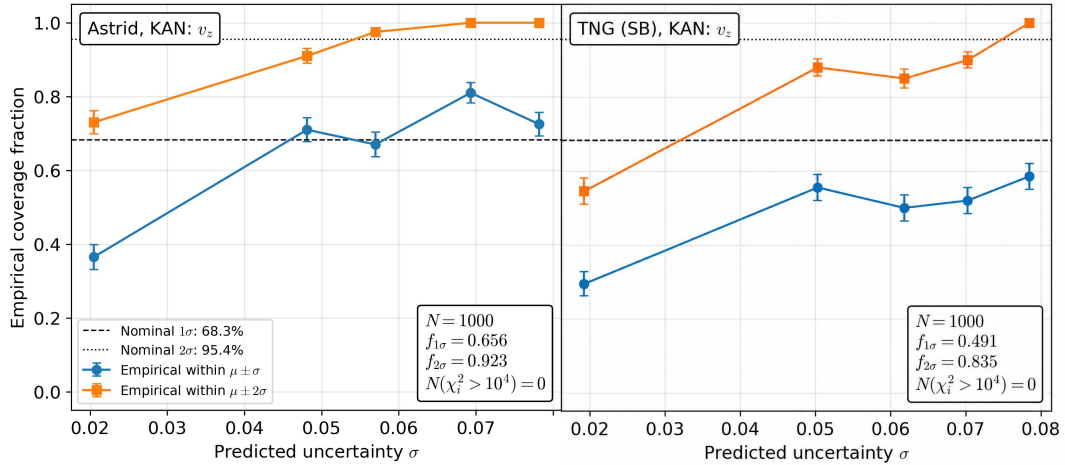}
    \caption{Empirical coverage of the predicted uncertainty estimates for the Astrid-trained DeepKAN model using line-of-sight peculiar velocities, $v_z$. Catalogs are binned by predicted uncertainty $\sigma_i$, and points show the fraction of true $\Omega_m$ values falling within $\mu_i\pm\sigma_i$ and $\mu_i\pm2\sigma_i$ in each bin. Dashed and dotted horizontal lines indicate the nominal Gaussian coverages of $68.3\%$ and $95.4\%$, respectively. The quoted $f_{1\sigma}$ and $f_{2\sigma}$ values are the aggregate coverage fractions over the full test set. The model has aggregate coverage close to the nominal values on held-out Astrid catalogs, with $f_{1\sigma}=0.656$ and $f_{2\sigma}=0.923$, while the cross-suite TNG(SB) catalogs show lower aggregate coverage, with $f_{1\sigma}=0.491$ and $f_{2\sigma}=0.835$}. In both cases, $N_{\rm clip}=0$.
    \label{fig:coverage}
\end{figure*}

Figure~\ref{fig:coverage} shows that the Astrid-trained DeepKAN model using $v_z$ has aggregate coverage close to the nominal Gaussian values on held-out Astrid catalogs, with aggregate coverage fractions $f_{1\sigma}=0.656$ and $f_{2\sigma}=0.923$. The binned curves show some variation with predicted uncertainty, so we interpret the figure primarily as an aggregate coverage diagnostic rather than as evidence for perfect conditional calibration. In the cross-suite TNG(SB) test, aggregate coverage decreases to $f_{1\sigma}=0.491$ and $f_{2\sigma}=0.835$, indicating that the predicted uncertainties become overconfident under distribution shift. No catalogs in either case satisfy $\chi_i^2>10^4$, so the observed undercoverage is not driven by the catastrophic outliers excluded from the clipped $\chi^2$ statistic. The elevated mean $\chi^2$ reported for the same model in Table~\ref{tab:feature_table} should therefore be interpreted together with the aggregate coverage fractions: the former is a squared-residual diagnostic that is sensitive to the high-$\chi_i^2$ tail, while the latter measures the fraction of catalogs contained within the predicted intervals.

Notably, despite variations in baryonic modeling across simulation suites, the correlation between predicted and true $\Omega_m$ remains strong, with PCC values of $\sim 0.75$--$0.8$ and relative errors of $\sim 25\%$ in cross-suite evaluation. This shows that peculiar velocities contain robust cosmological information and motivates the addition of positional information as input features to assess whether they provide additional constraining power for inferring $\Omega_m$. The lack of improvement with an increase in functional expressivity suggests that inference performance in this regime is primarily limited by information content rather than model capacity. 

This result contrasts with previous work using Deep Sets with MLPs, trained on Astrid, to infer $\Omega_m$ from line-of-sight peculiar velocities \citep{de_Santi_2023}. In their experiments, the authors found that line-of-sight velocities on their own were insufficient to extract a signal for $\Omega_m$ (see Figure~\ref{fig:metalayer}). Full details of the comparison with their work can be found in Appendix~\ref{appx:desanti}\footnote{This comparison depends on the specific layer architecture used to implement the set-based model. The contrast with \citet{de_Santi_2023} should therefore be regarded as provisional.}.

\subsection{Deep Sets with Spatial Information} \label{sec:ds_posits}

\begin{figure*}
    \centering
    \includegraphics[width=\linewidth]{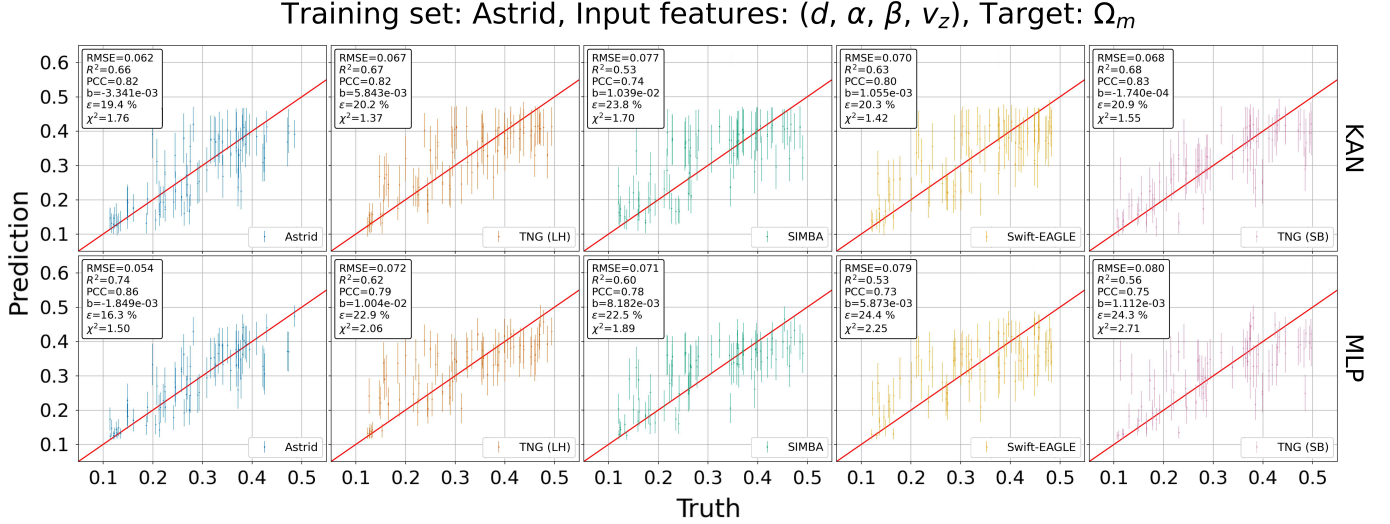}
    \caption{Comparing $\Omega_m$ inference using a DeepKAN model (top) and a Deep Set model implemented with multilayer perceptrons (MLPs) (bottom), using positions (transformed to be invariant to rotations and translations) and line-of-sight peculiar velocities ($v_z$) as input features. While performance on the training set degrades slightly, overall performance is comparable to the velocity-only model, highlighting the limitation of these set-based architectures to utilize relational structure to extract cosmological information.}
    \label{fig:kan_pos-vz_compare}
\end{figure*}

To assess whether additional spatial information improves inference within a permutation-invariant architecture, we trained both a DeepKAN and an MLP-based Deep Set model using transformed galaxy positions (see~\ref{sec:inputs}) together with line-of-sight peculiar velocities. Figure~\ref{fig:kan_pos-vz_compare} shows the corresponding results.

Across both architectures, adding positional information does not lead to consistent improvements in performance. In the DeepKAN case, positional information degrades in-distribution performance and yields mixed changes out-of-distribution, while MLP results remain similarly inconsistent. Overall, performance remains broadly comparable across feature sets. Correlation coefficients remain at the level PCC $\sim 0.82$--$0.86$, with relative errors of $\sim 20$--$24\%$ on OOD data, consistent with the results obtained using peculiar velocities alone.

\begin{figure*}
    \centering
    \includegraphics[width=\linewidth]{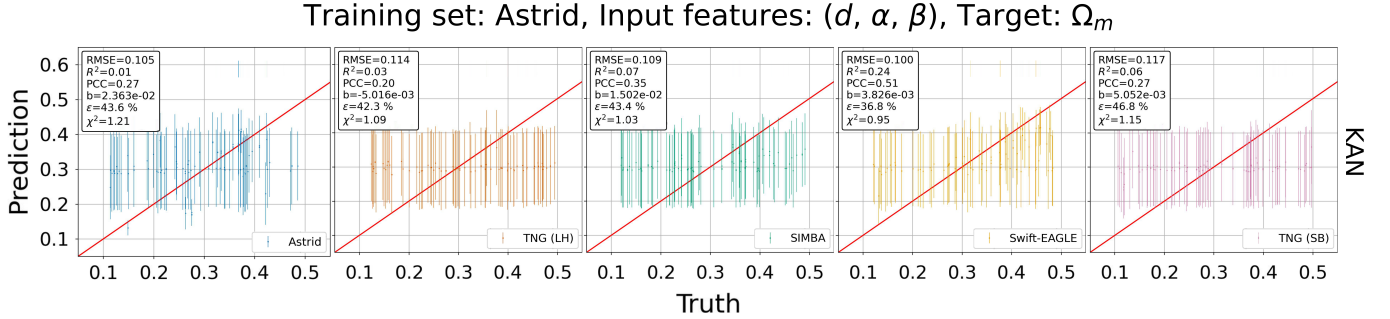}
    \caption{$\Omega_m$ inference using a DeepKAN model trained on the LH set of the Astrid simulation suite. Positions-only models show no recoverable cosmological signal (MLPs were also tested, with similar results), while the strong performance of the velocity-only and velocity+positions models indicates that, within permutation-invariant architectures, the dominant accessible signal arises from the velocity field. This suggests that Deep Sets can extract kinematic information effectively, but are unable to exploit spatial information without explicit relational structure.}
    \label{fig:kan_pos}
\end{figure*}

As an additional check on the information content of these models, we test them using only our transformed positions as input. Figure~\ref{fig:kan_pos} shows that positions by themselves carry no cosmological signal when evaluated using a permutation-invariant, set-based architecture. This indicates that, within a Deep Set framework, the dominant cosmological signal accessible to the network is already contained in the kinematic information. The addition of spatial coordinates does not significantly enhance performance, suggesting that positional information is not used effectively under global permutation-invariant aggregation. This result implies that the primary limitation of Deep Sets in this setting is not necessarily the absence of positional inputs, but the lack of an architectural mechanism to model relational structure between galaxies. This interpretation is also supported by the simple $\log N$ baseline in Table~\ref{tab:baselines}. The positions-only Deep Set achieves performance comparable to a model using only galaxy counts, indicating that this position parameterization provides little additional usable information to the set-based architecture beyond abundance information. Table~\ref{tab:feature_table} summarizes the results from all of our Deep Set experiments.

\subsection{GNNs and Spatial Information}

To test whether spatial information can be more effectively exploited by explicitly encoding relations between nearby galaxies, we trained graph neural networks (GNNs) on galaxy catalogs including both positions and line-of-sight peculiar velocities. In this framework, as laid out in subsection~\ref{sec:GNNs}, galaxies are represented as nodes and edges are constructed based on spatial proximity, enabling iterative message passing between neighboring systems. As mentioned in Section~\ref{sec:intro}, the use of GNNs for inferring $\Omega_m$ from a combination of simulated galaxy positions and other physical or kinematic features is well established and shown to be effective \citep{Villanueva_Domingo_2022, de_Santi_2023}. We present these results to explicitly highlight the differences between set-based and graph-based approaches in order to more clearly expose the effect of their respective inductive biases.

\begin{figure*}
    \centering
    \includegraphics[width=\linewidth]{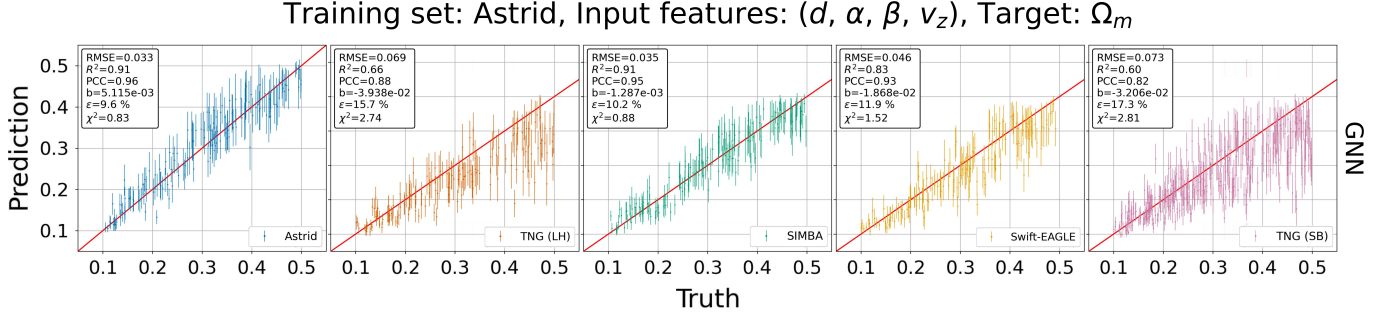}
    \caption{$\Omega_m$ inference using a GNN trained on the LH set of the Astrid simulation suite, using positions (transformed to be invariant to rotations and translations) and line-of-sight peculiar velocities ($v_z$) as input features. The model is tested on the LH sets of the IllustrisTNG, Simba, and Swift-EAGLE suites, as well as the SB28 set of IllustrisTNG. Graph-based architectures are able to take advantage of explicitly encoding relational information between nearby galaxies, resulting in improved inference on both in- and out-of-distribution suites.}
    \label{fig:gnns}
\end{figure*}

Figure~\ref{fig:gnns} shows the performance of the GNN models when trained on ASTRID and evaluated both in-distribution and across alternative simulation suites. Relative to permutation-invariant Deep Set architectures, we observe systematically improved predictive performance when positional information is included. In particular, cross-suite $R^2$ values increase and relative errors decrease, while the regression slope more closely follows the one-to-one relation.

This contrast with Section~\ref{sec:ds_posits} is significant. While raw spatial coordinates did not enhance inference within a Deep Set framework, they become informative when incorporated through local graph structure. This indicates that positional information contributes to cosmological inference primarily through relational patterns, such as clustering, rather than through individual galaxy coordinates treated independently.

The improved performance of GNNs therefore suggests that a substantial fraction of cosmological signal is encoded in the geometry of large-scale structure. Deep Sets, which aggregate over independent pointwise embeddings, cannot capture these relational dependencies. In contrast, GNNs introduce an architectural mechanism for modeling spatial relationships between galaxies, allowing positional information to be leveraged in a physically meaningful way.

Taken together, these results demonstrate that inference performance in this regime is limited not by function approximation capacity, but by the inductive biases imposed by the model architecture. When relational structure is incorporated, positional information becomes cosmologically informative.

\subsection{Failure of \texorpdfstring{$\sigma_8$}{sigma8} Inference}\label{sec:sig8_fail}

\begin{figure*}
\centering
\includegraphics[width=\linewidth]{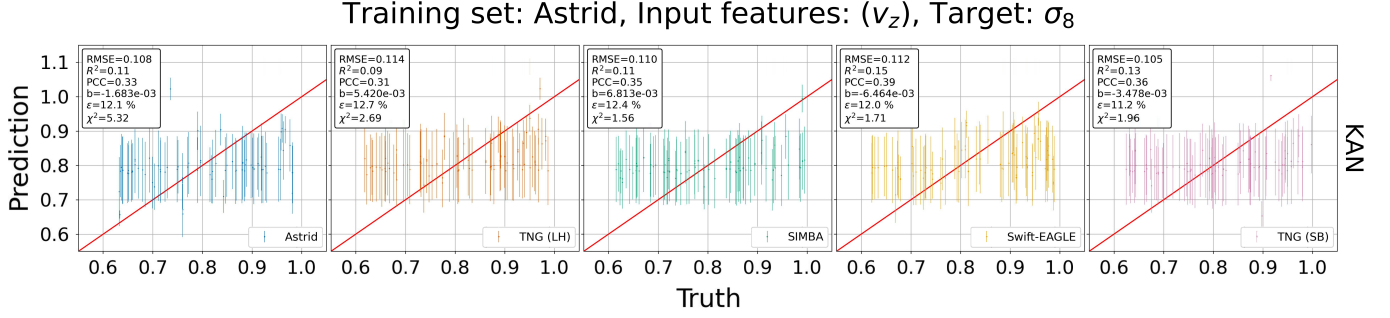}
\caption{$\sigma_8$ inference using a DeepKAN model trained on the LH set of the Astrid simulation suite, using only line-of-sight peculiar velocities as input. The model is tested on the LH sets of the Astrid, SIMBA, and Swift-EAGLE suites, as well as the SB28 set of IllustrisTNG. Unlike the corresponding $\Omega_m$ inference task, the model fails to recover a reliable signal for $\sigma_8$, consistent with previous field-level studies using galaxy catalogs \citep{Villanueva_Domingo_2022, de_Santi_2023}.}
\label{fig:kan_sig}
\end{figure*}

We also train the same DeepKAN architecture to infer $\sigma_8$ using only line-of-sight peculiar velocities. Figure~\ref{fig:kan_sig} shows the results. In contrast to the $\Omega_m$ case, the model fails to recover a reliable signal for $\sigma_8$ on both the Astrid test set and the cross-suite test sets. The predictions remain weakly correlated with the true values and do not provide useful parameter recovery.

This negative result is important for interpreting the cosmological information extracted by the model. The success of the velocity-only models for $\Omega_m$ does not imply that they extract all cosmological information present in the velocity field, nor that they break the usual amplitude-growth degeneracies associated with velocity statistics. Instead, the learned summaries appear to capture features of the line-of-sight velocity field that are informative about $\Omega_m$, while failing to isolate the fluctuation-amplitude information needed to constrain $\sigma_8$.

This behavior is consistent with previous field-level studies using galaxy catalogs. \cite{Villanueva_Domingo_2022} found that GNNs trained on galaxy positions and galaxy properties were unable to recover $\sigma_8$, and \cite{de_Santi_2023} similarly found that models using galaxy phase-space information did not robustly constrain $\sigma_8$. One plausible explanation is the limited $(25 h^{-1}\mathrm{Mpc})^3$ volume of the CAMELS boxes. Since $\sigma_8$ describes the amplitude of matter fluctuations on $8h^{-1}\mathrm{Mpc}$ scales, each catalog samples only a small number of independent regions on the relevant scale and lacks longer-wavelength modes that contribute to coherent velocity flows. The failure to recover $\sigma_8$ should therefore be understood as evidence that, in this small-volume setting, the learned velocity summaries retain useful information about $\Omega_m$ but do not provide a recoverable constraint on the fluctuation amplitude. We return to this point in Section~\ref{sec:velos_om_sig8} in light of the linear-theory dependence of the velocity power spectrum on $\sigma_8$.

\section{Discussion}
\label{sec:discussion}

\subsection{Peculiar Velocities, $\Omega_m$, and $\sigma_8$}\label{sec:velos_om_sig8}

Peculiar velocities provide a direct probe of the growth of cosmic structure and therefore of the underlying matter density parameter $\Omega_m$ \citep{Hudson_2012, Turner2025, Qin_2021}. While galaxy positions trace the spatial distribution of matter, peculiar velocities respond directly to gradients in the gravitational potential and consequently to the clustering of matter itself. Because gravitational instability governs the growth of large-scale structure, the magnitudes and spatial coherence of peculiar velocities encode information about the matter density that drives structure formation \citep{Turner2025}. 
The relevant question is therefore not whether peculiar velocities contain cosmological information in principle, but which parameter dependencies remain recoverable from the finite, nonlinear galaxy catalogs used here.

This connection originates in the Newtonian fluid description of structure formation. In comoving coordinates, the Poisson equation,

\begin{equation}\label{eq:poisson}
    \nabla^2\Phi = \frac{3}{2}\Omega_m H^2 a^2 \delta.
\end{equation}

\noindent where $a$ is the scale factor, $H$ is the Hubble parameter, and $\delta=(\rho-\bar{\rho})/\bar{\rho}$ is the density contrast, relates the gravitational potential $\Phi$ to the matter overdensity at all scales, with $\Omega_m$ controlling the strength of the gravitational source term. Peculiar velocities arise from the gradient of this potential via the Euler equation, 

\begin{equation}\label{eq:euler}
    \frac{\partial \mathbf{v}}{\partial t}
    + H \mathbf{v}
    + \frac{1}{a}(\mathbf{v}\cdot\nabla)\mathbf{v}
    = -\frac{1}{a}\nabla\Phi,
\end{equation}

\noindent so $\Omega_m$ sets the normalization of the velocity field regardless of whether the density field is in the linear or non-linear regime. The continuity equation,

\begin{equation}\label{eq:continuity}
    \frac{\partial \delta}{\partial t}
    + \frac{1}{a}\nabla\cdot\left[(1+\delta)\mathbf{v}\right] = 0,
\end{equation}

\noindent further links the velocity divergence to the time evolution of the density field. Together, these equations form a closed dynamical system (see, e.g., \cite{Peebles_1980_LSS, Bernardeau_2002, Turner2025}) in which $\Omega_m$ governs how efficiently density perturbations generate coherent motions.

In the linear regime, these equations simplify, yielding a direct relation between the velocity divergence and the matter overdensity,
\begin{equation}
    \nabla \cdot \mathbf{v}(\mathbf{r},z)
        = -\,a\,H\,f(z)\;\delta(\mathbf{r},z),
\end{equation}

\noindent where $f(z)\equiv d\ln D/d\ln a$ is the linear growth rate. In $\Lambda$CDM, this is well approximated by $f(z)\simeq \Omega_m(z)^{\gamma}$ with $\gamma \approx 0.55$ \citep{PhysRevD.72.043529}, providing the leading-order sensitivity of peculiar velocities to $\Omega_m$. This linear relation implies that the velocity field is Gaussian and fully characterized by its power spectrum \citep{Peebles_1993}, 

\begin{equation}\label{eq:vpower}
    P_{vv}(k,z) = \left(\frac{aHf}{k}\right)^{2} P_{\delta\delta}(k,z),
\end{equation}

\noindent which depends on the combination $f\sigma_8$, since the normalization of $P_{\delta\delta}\propto\sigma_8^2$.

Equation~\ref{eq:vpower} also shows why the failure to recover $\sigma_8$ in Section~\ref{sec:sig8_fail} is informative. Since $\sigma_8$ enters the linear velocity power spectrum through the amplitude of $P_{\delta\delta}$, the null result should not be interpreted as evidence that the velocity field is independent of $\sigma_8$. Instead, it indicates that the fluctuation-amplitude information is not recoverable by the learned summaries in this particular setup. This is plausibly dominated by the small simulation volume: a $(25h^{-1}\mathrm{Mpc})^3$ box contains only a limited sampling of the $8h^{-1}\mathrm{Mpc}$ scale associated with $\sigma_8$ and omits longer-wavelength modes that contribute to coherent velocities. In contrast, the successful recovery of $\Omega_m$ indicates that some information tied to the gravitational sourcing and growth of the velocity field remains accessible even on the nonlinear scales probed by these catalogs.

The $\Omega_m$ side of this contrast can be understood from the dynamical connection that persists beyond the linear regime. On the non-linear scales probed by the simulations used in this work, gravitational evolution departs from linear theory, inducing mode coupling and generating non-Gaussian features in the velocity field \citep{Bernardeau_2002}. For the velocity-only Deep Set models considered here, this motivates asking whether the empirical distribution of line-of-sight velocities contains cosmological information beyond simple hand-crafted one-point summaries. The fundamental connection between peculiar velocities and $\Omega_m$ persists via the Poisson equation (Eq~\ref{eq:poisson}), which ensures that $\Omega_m$ controls the gravitational sourcing of velocities in the Newtonian regime. Moreover, linear theory implies that velocity fields are more strongly weighted toward large scales than density fields. This helps explain why velocity statistics are often less affected by small-scale nonlinear structure, and can therefore remain closer to linear-theory expectations than density statistics, even though the exact linear relations break down once the field becomes fully nonlinear \citep{Jennings_2011, Koda_2014}. The $\Omega_m$ results show, then, that the learned summaries capture useful information in the empirical distribution of line-of-sight velocities beyond simple hand-crafted one-point summaries, while the $\sigma_8$ failure shows that this information is not sufficient to recover all parameter dependencies present in principle in the velocity field. This statement is narrower than the information-content comparisons made in explicit likelihood-based field-level analyses, which can test the gain over optimized two- and three-point clustering statistics using controlled forward models \citep{Nguyen_2024_FieldLevelInfo, Akitsu_2025_FieldLevelBispectrum}.

Although simulations provide the full three-dimensional velocity field, this work uses only one projected component, $v_{\parallel}=\mathbf{v}\cdot\hat{\mathbf{n}}$, motivated by the line-of-sight nature of observed peculiar velocities. Under statistical isotropy, this projected component is governed by the same gravitational dynamics as the full velocity field and therefore retains information about the amplitude and coherence of cosmic velocity flows. The successful recovery of $\Omega_m$ from simulated $v_z$ demonstrates that this projected velocity information is sufficient for the specific inference task considered here, while the failure to recover $\sigma_8$ shows that it is not sufficient for all growth-related parameters in this finite-volume setting.

\subsection{Information Content and Architectural Inductive Bias}

The results of the Deep Set experiments presented in Section~\ref{sec:results} demonstrate that adding positional information does not substantially improve inference when using a permutation-invariant architecture. This does not imply that positions lack cosmological information. Rather, it indicates that spatial information is not effectively utilized when treated as independent pointwise features aggregated globally.

Clustering information is inherently relational, arising from correlations between the positions of galaxies rather than from their individual properties. In particular, cosmological information encoded in spatial structure depends on pairwise and higher-order relationships, such as separations and local environments \citep{2005MNRAS.357..608Y, Hamilton_1998}, which cannot be reduced to independent pointwise features. Architectures that explicitly model the spatial relationships between galaxies, such as GNNs, are therefore naturally suited to extracting this information. In contrast, Deep Set architectures operate on individual galaxy embeddings and aggregate them globally, without representing explicit relationships between galaxies. As a result, positional information is not efficiently utilized in a set-based framework, and the dominant cosmological signal captured by such models is instead contained in the kinematic information of peculiar velocities.

The improved performance of GNNs when positional information is included is therefore not unexpected. GNNs introduce relational inductive bias through message passing along edges defined by spatial proximity \citep{gilmer2017neuralmessagepassingquantum}. This mechanism allows the model to capture pairwise and higher-order correlations analogous to those encoded in traditional clustering statistics. The enhancement in performance observed with GNNs suggests that positional information becomes cosmologically informative when relational structure is explicitly modeled. At the same time, these architectural comparisons should be interpreted conditional on the specific implementations and hyperparameter searches used in this work. The layer architecture discrepancy discussed in Appendix~\ref{appx:desanti} shows that nominally similar set-based implementations can differ in practice, so the cross-suite degradation pattern, the GNN improvement over Deep Sets, and the KAN-MLP equivalence should be understood as empirical results for the models tested here rather than implementation-independent guarantees.

Taken together, the results indicate that inference performance in this regime is constrained not by function approximation capacity, but by architectural inductive bias. Increasing expressivity within a permutation-invariant framework, such as replacing MLP-based Deep Sets with DeepKANs, does not lead to substantial improvement, indicating that the limiting factor is not the flexibility of the learned functions. In contrast, incorporating relational structure does improve performance, but only when paired with an architecture capable of exploiting it. Graph neural networks, which explicitly model spatial structure between galaxies, are able to extract additional geometric information from the galaxy distribution, whereas set-based architectures are not. The empirical coverage test further indicates that the predicted heteroscedastic uncertainties are informative but become overconfident under cross-suite distribution shift, so robustness should be interpreted in terms of both point-prediction accuracy and uncertainty calibration.

An important caveat in interpreting our results is that the present analysis uses exact simulated line-of-sight peculiar velocities, whereas in real surveys peculiar velocities are not directly observed. Instead, they are inferred by combining spectroscopic redshifts with redshift-independent distance indicators such as the Fundamental Plane \citep{Djorgovski1987}, Tully--Fisher relation \citep{TullyFisher1977}, Faber--Jackson relation \citep{FaberJackson1976}, or Type Ia supernovae \citep{Branch1993}. As a result, observational peculiar-velocity catalogs are affected by distance uncertainties, calibration systematics, selection effects, and incomplete sky coverage \citep{Scrimgeour_2014_6dFGSv, Tully_2023_Cosmicflows4}. The results presented here should therefore be understood as demonstrating the cosmological information content available in an idealized peculiar-velocity observable, rather than as a direct forecast for any specific survey.

\section{Conclusion} \label{sec:conclusions}

In this work, we investigated how architectural inductive bias and functional expressivity affect cosmological parameter inference from simulated galaxy catalogs. Focusing on $\Omega_m$ as a representative growth-sensitive parameter, we compared permutation-invariant Deep Set models, implemented with both standard MLPs and a relatively new functional approximator called Kolmogorov–Arnold Networks (KANs), as well as relational graph neural networks (GNNs), using peculiar velocities and spatial information as input features.

Our results suggest that field-level machine learning provides a flexible alternative to traditional summary-statistic pipelines. Rather than explicitly constructing correlation functions or multipoles, the models learn mappings directly from galaxy catalogs to cosmological parameters. We also demonstrate that line-of-sight peculiar velocities alone contain substantial cosmological information. Velocity-only Deep Set models recover $\Omega_m$ with competitive performance both in-distribution and across simulation suites with differing baryonic prescriptions. This finding is consistent with the theoretical expectation that peculiar velocities directly trace the growth of structure through large-scale gravitational flows \citep{Feldman:2003xr, Nusser_2011}. However, this success does not extend to all cosmological parameters. In the same setting, the models do not recover a robust signal for $\sigma_8$, indicating that the velocity-based information extracted here is more effective for constraining $\Omega_m$ than the fluctuation-amplitude parameter.

Despite their alternative finite-size parameterization, KAN-based components do not yield significant improvement in inference accuracy in this regime. This suggests that performance is not limited by the flexibility of univariate function approximation, but rather by the information content of the input features and the inductive biases imposed by the architecture.

In contrast, when positional information is incorporated within a relational framework, GNNs exhibit improved performance relative to Deep Set models. This enhancement reflects the geometric nature of clustering statistics, which depend on spatial relationships and anisotropic correlations that are not explicitly modeled by permutation-invariant architectures. The results therefore highlight the importance of aligning architectural design with the physical structure of the underlying cosmological fields.

Taken together, these findings suggest a hierarchy of considerations for field-level cosmological inference. First, peculiar velocities provide a physically motivated and growth-sensitive probe that can be effectively exploited within set-based models. Second, architectural inductive bias is more consequential than the choice of pointwise function approximator when extracting cosmological information from structured data. Finally, relational modeling becomes essential when spatial geometry is expected to carry additional signal.

An important limitation of this analysis is that it uses exact simulated line-of-sight peculiar velocities, whereas real peculiar-velocity catalogs require distances inferred from noisy, heterogeneous observational indicators and are affected by calibration systematics, selection effects, and survey geometry. As forthcoming surveys substantially expand the volume and precision of peculiar velocity measurements, inference frameworks that respect both the dynamical and geometric structure of large-scale structure will play an increasingly important role. The implications of these results for survey pipelines are therefore contingent on validation under realistic velocity-measurement noise, selection effects, and survey geometry. By interpreting machine learning architectures through the lens of cosmological theory, we can better understand not only how well models perform, but why they succeed or fail in capturing the information encoded in the cosmic web.

\section*{Acknowledgments} \label{sec:ack}
We are grateful to Natal\'i de Santi for discussions and insight into previous work and for sharing her code so that accurate and reliable comparisons could be carried out. The CAMELS project is supported by the Simons Foundation and NSF grant AST 2108078. The work of S.G. and F.V.N. is supported by the Simons Foundation. The training of the Deep Sets and GNNs has been carried out using graphics processing units (GPUs) from Simons Foundation, Flatiron Institute, Center for Computational Astrophysics.

\newpage
\bibliography{refs}{}
\bibliographystyle{aasjournalv7}



\appendix

\section{trained on TNG}

\begin{figure*}
    \centering
    \includegraphics[width=\linewidth]{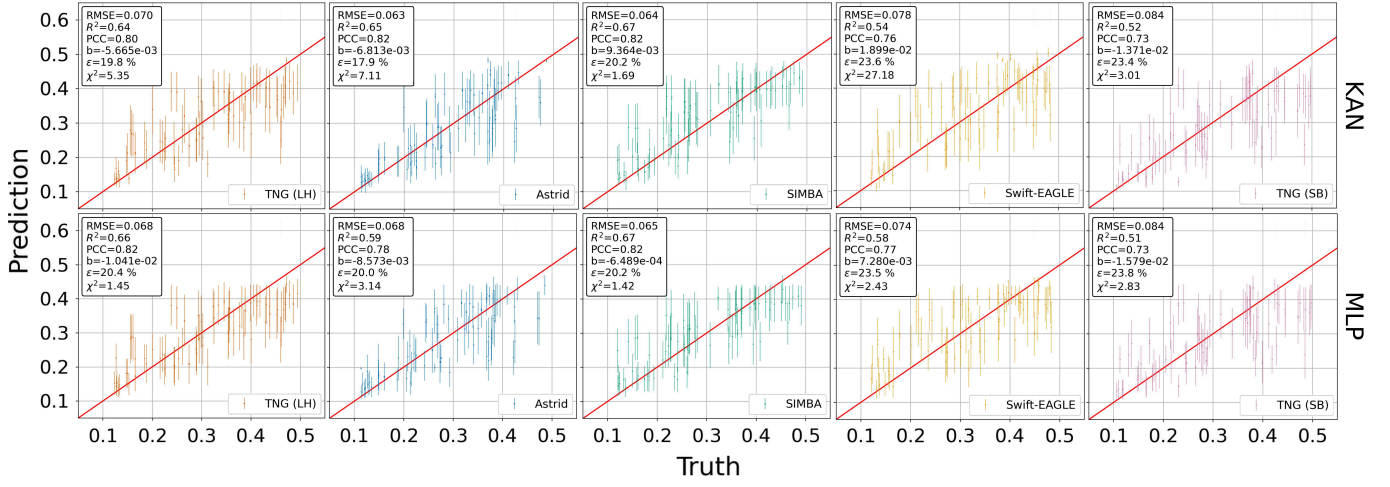}
    \caption{Comparing $\Omega_m$ inference between a DeepKAN model (top) and a Deep Set implemented with multilayer perceptrons (MLPs) (bottom), trained on the LH set of the IllustrisTNG simulation suite, using only the line-of-sight peculiar velocities as an input feature. The comparable performance between the two models, and the comparable performance with models trained on Astrid, suggests that the cosmological information accessible through the velocity field is relatively robust for the LH sets of the Astrid and TNG simulations.}
    \label{fig:tng_vz_compare}
\end{figure*}

We train our Deep Set with MLPs and our DeepKAN using only the line-of-sight peculiar velocities on the LH set of IllustrisTNG, and test the model on Astrid, Simba, Swift-EAGLE, and the SB set of IllustrisTNG. The model is optimized using the same Optuna framework as the model trained on Astrid, and trained over 300 epochs using the best performing hyperparameters. The results are shown in Figure~\ref{fig:tng_vz_compare} and summarized in Table~\ref{tab:tng_vz_comparison}. While the in-distribution performance of models trained on TNG (LH) is somewhat degraded relative to models trained on Astrid, the overall behavior is broadly consistent across both training suites and across both Deep Sets with MLPs and Deep KANs. This suggests that for at least these two simulation suites, the velocity field retains cosmological information that can be learned by permutation-invariant, set-based architectures. At the same time, these results indicate that the successful recovery of this information depends on the training data provided and therefore should not be interpreted as evidence of suite-independent robustness, in general.

\begin{table*}[t]
\centering
\begin{tabularx}{0.8\textwidth}{llccccccc}
\hline
\textbf{Test Suite} & \textbf{Model} & \textbf{RMSE} & \textbf{$R^2$} & \textbf{PCC} & \textbf{Bias} & \textbf{$\epsilon$ (\%)} & \textbf{$\chi^2$} & \textbf{$N_{\rm clip}$} \\
\hline

\multirow{2}{*}{\textbf{TNG (LH)}}
& Deep Set (MLP) & 0.068 & 0.66 & 0.82 & $-1.04\times10^{-2}$ & 20.4 & 1.45 & 0 \\
& DeepKAN        & 0.070 & 0.64 & 0.80 & $-5.67\times10^{-3}$ & 19.8 & 5.35 & 0 \\
\hline

\multirow{2}{*}{\textbf{ASTRID}}
& Deep Set (MLP) & 0.068 & 0.59 & 0.78 & $-8.57\times10^{-3}$ & 20.3 & 3.14 & 0 \\
& DeepKAN        & 0.063 & 0.65 & 0.82 & $-6.81\times10^{-3}$ & 17.9 & 7.11 & 0 \\
\hline

\multirow{2}{*}{\textbf{SIMBA}}
& Deep Set (MLP) & 0.065 & 0.67 & 0.82 & $-6.49\times10^{-4}$ & 20.2 & 1.42 & 0 \\
& DeepKAN        & 0.064 & 0.67 & 0.82 & $9.36\times10^{-3}$ & 20.2 & 1.69 & 0 \\
\hline

\multirow{2}{*}{\textbf{Swift-EAGLE}}
& Deep Set (MLP) & 0.074 & 0.58 & 0.77 & $7.28\times10^{-3}$ & 23.5 & 2.43 & 0 \\
& DeepKAN        & 0.078 & 0.54 & 0.76 & $1.89\times10^{-2}$ & 23.6 & 27.19 & 2 \\
\hline

\multirow{2}{*}{\textbf{TNG (SB)}}
& Deep Set (MLP) & 0.084 & 0.51 & 0.73 & $-1.58\times10^{-2}$ & 23.8 & 2.83 & 0 \\
& DeepKAN        & 0.084 & 0.52 & 0.73 & $-1.37\times10^{-2}$ & 23.4 & 3.01 & 0 \\
\hline

\end{tabularx}
\caption{Comparison of Deep Set models for inferring $\Omega_m$ using only line-of-sight peculiar velocities ($v_z$). All models are trained on the LH set of IllustrisTNG and evaluated on multiple simulation suites. Metrics reported are RMSE, $R^2$, PCC, relative error $\epsilon$, and reduced clipped $\chi^2$. Here $\chi^2$ is computed after excluding catalogs with $\chi^2>10^4$, and $N_{\rm clip}$ gives the number of excluded catalogs. The global feature $u=\text{log}N$ is included in all training and test samples.}
\label{tab:tng_vz_comparison}
\end{table*}

\section{Comparison with previous work} \label{appx:desanti}

In order to draw a comparison between DeepKANs and a Deep Set with MLPs, we sought to reproduce the results from \cite{de_Santi_2023}. In that work, the authors build on experiments carried out in \cite{Villanueva_Domingo_2022}, where simulated galaxy catalogs from the IllustrisTNG and SIMBA suites are analyzed using a GNN. All experiments are implemented within the \textsc{CosmoGraphNet} framework \citep{Villanueva_Domingo_2022, pablo_villanueva_domingo_2022_6485804}\footnote{The GitHub repository is available at \href{https://github.com/PabloVD/CosmoGraphNet/tree/master?tab=readme-ov-file}{this link}.}, which provides a pipeline for cosmological inference using GNN architectures. To isolate the information content of peculiar velocities, \cite{de_Santi_2023} remove all edge connections from the graphs, eliminating message passing between nodes. In this configuration, each galaxy is processed independently by a shared neural network, followed by a permutation-invariant aggregation over nodes. This architecture is therefore functionally equivalent to a Deep Set model, with the implementation carried out using the \texttt{MetaLayer} class from the \texttt{PyTorch Geometric (PyG)} library \citep{fey2019fastgraphrepresentationlearning} to construct stacks of MLPs. In contrast, our implementation of Deep Sets is built directly using standard PyTorch modules, with fully connected layers implemented via the \texttt{Linear} class for MLP-based models \citep{Paszke_2019_PyTorch}, and replaced by KAN layers in the DeepKAN variant.

\begin{figure*}
    \centering
    \includegraphics[width=\linewidth]{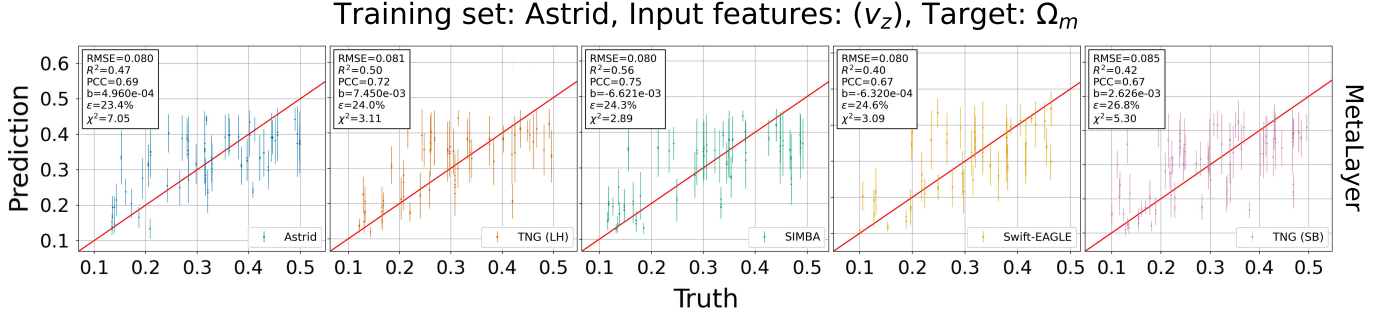}
    \caption{Model trained using the \texttt{MetaLayer} class from the \texttt{PyG} library, in keeping with the approach of \cite{de_Santi_2023}. Comparing this to Figure~\ref{fig:kan_vz_compare}, we see that the model fails to recover a significant cosmological signal. The specific reasons that the \texttt{MetaLayer} architecture underperforms when compared to PyTorch's \texttt{Linear} class are not clear.}
    \label{fig:metalayer}
\end{figure*}

Results from performing inference using the \texttt{PyG} implementation, and a table summarizing those results in contrast to our own, can be found in Figure~\ref{fig:metalayer} and Table~\ref{tab:desanti_table}, respectively. It is evident that the \texttt{MetaLayer} architecture struggles to extract the underlying $\Omega_m$ signal. The discrepancy is therefore localized to the layer architecture rather than to the input features or training suite. However, the mechanism by which the \texttt{Linear}-based and \texttt{MetaLayer}-based implementations lead to different optimization behavior and predictive performance is not yet fully understood. The comparison with \citet{de_Santi_2023} should therefore be regarded as provisional pending a more detailed diagnostic study of this implementation dependence.

Additionally, the \texttt{MetaLayer} results reported here differ from those of \citet{de_Santi_2023}. For the model trained and tested on Astrid, \citet{de_Santi_2023} report $R^2\sim0.2$, with OOD values averaging $\sim0.25$. In contrast, we find $R^2=0.47$ in distribution and an average OOD value of $R^2\sim0.47$. This difference likely reflects the hyperparameter optimization strategy. \citet{de_Santi_2023} used 20 Optuna trials of 300 epochs each, whereas we used 100 trials of 20 epochs each. In our experiments, 20 epochs were sufficient to identify promising configurations, and the larger number of trials allowed Optuna to explore the hyperparameter space more broadly and select a more effective model. 

\begin{table*}[t]
\centering
\begin{tabularx}{0.835\textwidth}{llcccccc}
\hline
\textbf{Test Suite} & \textbf{Model} & \textbf{RMSE} & \textbf{$R^2$} & \textbf{PCC} & \textbf{Bias} & \textbf{$\epsilon$ (\%)} & \textbf{$\chi^2$} \\
\hline

\multirow{3}{*}{\textbf{ASTRID}}
& Deep Set (KAN)        & 0.057 & 0.71 & 0.85 & $1.28\times10^{-2}$ & 18.6 & 1.98 \\
& Deep Set (MLP, Linear)   & 0.052 & 0.76 & 0.87 & $5.51\times10^{-3}$ & 17.1 & 0.87 \\
& Deep Set (MLP, MetaLayer) & 0.080 & 0.47 & 0.69 & $4.95\times10^{-4}$ & 23.4 & 7.05 \\
\hline

\multirow{3}{*}{\textbf{TNG (LH)}}
& Deep Set (KAN)        & 0.076 & 0.58 & 0.77 & $1.67\times10^{-2}$ & 25.3 & 1.64 \\
& Deep Set (MLP, Linear)   & 0.073 & 0.61 & 0.79 & $1.45\times10^{-2}$ & 24.3 & 1.27 \\
& Deep Set (MLP, MetaLayer) & 0.081 & 0.51 & 0.72 & $7.44\times10^{-3}$ & 24.0 & 3.11 \\
\hline

\multirow{3}{*}{\textbf{TNG (SB)}}
& Deep Set (KAN)        & 0.080 & 0.56 & 0.76 & $1.32\times10^{-2}$ & 25.6 & 5.79 \\
& Deep Set (MLP, Linear)   & 0.080 & 0.56 & 0.75 & $8.20\times10^{-3}$ & 26.2 & 1.65 \\
& Deep Set (MLP, MetaLayer) & 0.085 & 0.42 & 0.67 & $2.62\times10^{-3}$ & 26.8 & 5.30 \\
\hline

\multirow{3}{*}{\textbf{SIMBA}}
& Deep Set (KAN)        & 0.074 & 0.57 & 0.77 & $1.81\times10^{-2}$ & 24.8 & 2.54 \\
& Deep Set (MLP, Linear)   & 0.073 & 0.58 & 0.78 & $1.39\times10^{-2}$ & 24.5 & 1.18 \\
& Deep Set (MLP, MetaLayer) & 0.080 & 0.56 & 0.75 & $6.61\times10^{-3}$ & 24.2 & 2.89 \\
\hline

\multirow{3}{*}{\textbf{Swift-EAGLE}}
& Deep Set (KAN)        & 0.077 & 0.55 & 0.77 & $2.38\times10^{-2}$ & 25.4 & 15.76 \\
& Deep Set (MLP, Linear)   & 0.081 & 0.51 & 0.72 & $1.10\times10^{-2}$ & 25.8 & 1.38 \\
& Deep Set (MLP, MetaLayer) & 0.080 & 0.40 & 0.67 & $-6.31\times10^{-4}$ & 24.5 & 3.09 \\
\hline

\end{tabularx}
\caption{Comparison of Deep Set models for inferring $\Omega_m$ from peculiar velocities. All models are trained on ASTRID, utilizing the \textsc{CosmoGraphNet} framework presented in \cite{Villanueva_Domingo_2022}, and evaluated on multiple simulation suites. Metrics reported are RMSE, $R^2$, PCC, relative error $\epsilon$, and reduced $\chi^2$.}
\label{tab:desanti_table}
\end{table*}

\end{document}